\documentclass[final,5p,times,twocolumn]{elsarticle}

\usepackage{multicol}
\usepackage[export]{adjustbox}
\usepackage{tabularray}
\usepackage{longtable}
\usepackage{booktabs}
\usepackage{amssymb}
\usepackage{subcaption}
\usepackage{threeparttable}
\usepackage{tabularx}
\usepackage{makecell}
\usepackage{pifont} 
\newcommand{\cmark}{\ding{51}} 
\newcommand{\xmark}{\ding{55}}%

\usepackage{xcolor}

\usepackage{multirow}
\usepackage{url}

\usepackage{graphicx}
\usepackage{underscore}
\usepackage[english]{babel}
\usepackage{lineno} 

\usepackage{svg}
\DeclareUnicodeCharacter{2212}{\ensuremath{-}}

\journal{Journal of Applied Acoustics}

\begin{document}

\begin{frontmatter}



\title{{A Data-Driven Exploration of Elevation Cues in HRTFs: An Explainable AI Perspective Across Multiple Datasets}}

\author[label-UV]{Juan~A.~De Rus\corref{correspondingauthor}}
\cortext[correspondingauthor]{Corresponding author}
\ead{juan.rus@uv.es}
\author[label-UV,label-i2CAT]{Mario Montagud}
\author[label-UV]{Jesus~Lopez-Ballester}
\author[label-UV]{Francesc J. Ferri}
\author[label-UV]{Maximo~Cobos}

\address[label-UV]{Departament d'Inform\`atica, Universitat de Val\`encia, 46100 Burjassot, Spain}
\address[label-i2CAT]{i2CAT Foundation, 08034 Barcelona, Spain}

\begin{abstract}
Precise elevation perception in binaural audio remains a challenge, despite extensive research on head-related transfer functions (HRTFs) and spectral cues. While prior studies have advanced our understanding of sound localization cues, the interplay between spectral features and elevation perception is still not fully understood. This paper presents a comprehensive analysis of over 600 subjects from 11 diverse public HRTF datasets, employing a convolutional neural network (CNN) model combined with explainable artificial intelligence (XAI) techniques to investigate elevation cues. In addition to testing various HRTF pre-processing methods, we focus on both within-dataset and inter-dataset generalization and explainability, assessing the model's robustness across different HRTF variations stemming from subjects and measurement setups. By leveraging class activation mapping (CAM) saliency maps, we identify key frequency bands that may contribute to elevation perception, providing deeper insights into the spectral features that drive elevation-specific classification. This study offers new perspectives on HRTF modeling and elevation perception by analyzing diverse datasets and pre-processing techniques, expanding our understanding of these cues across a wide range of conditions.\end{abstract}



\begin{keyword}
HRTF\sep spatial audio\sep elevation cues\sep convolutional neural networks\sep explainable artificial intelligence.

\end{keyword}

\end{frontmatter}


\section{Introduction}
Head-related impulse responses (HRIRs)  describe how sound is transformed as it travels from a source to the ears of a listener, capturing acoustic cues that depend on the individual’s unique anthropometry, such as head shape, ear structure, and torso \cite{blauert1997spatial}. In the frequency domain, these time-domain responses are represented as head-related transfer functions (HRTFs), which describe the frequency-dependent filtering characteristics of the torso, head, and pinnae \cite{Li2020}. HRTFs play a crucial role in sound localization, \cite{carlini2024auditory} where interaural level and time differences assist in left-right positioning, while monaural spectral cues are essential for determining up-down and front-back localization \cite{carlile1994location, Ahveninen2014}. Despite the challenges posed by the highly individual nature of spectral cues, understanding and analyzing HRTFs is essential for crafting immersive audio experiences in virtual reality, gaming, and realistic spatial audio systems \cite{Geronazzo2018, andersen2021evaluation, Rajendran2019}, where accurately conveying the intended elevation to users remains a key challenge.

Spectral cues arise from the acoustic filtering effects of the auditory periphery, including the torso, shoulders, head, and pinnae. The pinnae, with its complex structure of cavities and folds, is particularly important in producing resonances and reflections that influence sound localization \cite{Hebrank1974, Butler1977, Alves-Pinto2014, yao2020role, Stevenson2022}. While the role of the pinnae in localization is well established, the precise mechanisms governing elevation perception have been the focus of research since the late 1960s \cite{Shaw1968}. Spectral cues operate over a wide frequency range, particularly between 3 kHz and 9 kHz. Hebrank and Wright \cite{Hebrank1974}, along with Langendijk and Bronkhorst \cite{Langendijk2002}, identified a 1-octave notch in the 5-8 kHz range and a prominent peak around 13 kHz as key cues for frontal localization, with the latter also playing a role in rear HRTFs. A peak between 7 and 9 kHz, along with a high-frequency cut-off near 10 kHz, is associated with higher elevations. Furthermore, studies have shown that increases in frontal elevation correspond with an upward shift in the center frequency of the 1-octave notch, from approximately 5 to 11 kHz \cite{Hebrank1974, shaw1997acoustical, Langendijk2002}.

Recent research has continued to explore these dynamics, providing deeper insights into the frequency ranges relevant for elevation cues. Yao et al. \cite{yao2020role} identified three critical frequency bands (400 Hz to 1.2 kHz, 4 to 8 kHz, and 12 to 14 kHz) through a high Fisher’s F-ratio, with perceptual experiments revealing that variations in these bands significantly impact vertical localization.
Some works have already observed an impact of low frequencies in the perception of elevation \cite{Algazi2001ElevationSpectralCuesLowFreq, asano1990role, morimoto2003role} and according to Zonooz et al. \cite{zonooz2018learning, Zonooz2019}, the auditory system may employ a weighted spectral analysis to estimate source elevation. Additionally, studies by Reiss and Young \cite{reiss2005spectral} have highlighted that neural circuits involved in sound localization are highly sensitive to steep spectral slopes, akin to how the visual system processes edges and boundaries. In this context, the spectral derivative cue, introduced in \cite{zakarauskas1993computational} and later refined in \cite{baumgartner2014modeling}, draws on neurophysiological evidence suggesting that mammals decode spatial cues by focusing on spectral gradients, apart from center frequencies of peaks and notches. 

Identifying relevant localization cues is essential for developing binaural localization models that predict the direction of incoming sound. These models help explain auditory perception and evaluate the spatial quality of audio systems. Notable models include those of Langendijk et al. \cite{Langendijk2002} and Baumgartner et al. \cite{Baumgartner2013}. While acknowledging the role of many of the above effects, existent models do not fully explain how the brain extracts common elevation cues that are consistent across individuals, despite variations in pinnae shapes. 

As a result, accurately conveying vertical auditory perception in binaural audio remains a challenge, underscoring the need for further exploration to improve the reproduction of elevation cues in virtual sound environments.

Numerous approaches have been employed to address these challenges by modeling HRTFs. Geometrical models, like spherical head models and boundary element methods (BEM) \cite{algazi2001estimation, Algazi2002AproximatingHRTFGeomModels, Xiao2003, katz2001boundary}, simulate the acoustic effects of head, ear, and torso shapes on sound propagation. Principal component analysis (PCA) \cite{Liang2009} has efficiently captured individual HRTF variations, offering compact representations of spatial hearing. Cylindrical and spherical harmonics have been used to efficiently represent HRTFs by modeling them as a sum of multiple spatial components \cite{pollow2012calculation,Zotkin2009}. This approach provides a compact and continuous description of spatial cues, simplifying the modeling and computation of HRTFs across various spatial positions while reducing complexity. Recently, machine learning and deep learning techniques have introduced data-driven methods for modeling and predicting personalized HRTFs based on anthropometric features \cite{Zhu2017,Lee2018, Fantini21, Zhao22}. These diverse methodologies collectively highlight the multidimensional efforts made in understanding and simulating HRTFs for accurate spatial audio perception. Nonetheless, despite advancements in HRTF modeling and processing, the full comprehension and faithful synthesis of elevation cues remain elusive. Among the mentioned techniques, the utilization of deep neural networks (DNNs) for HRTF processing has been steadily on the rise \cite{cobos2022overview}. Recent studies have demonstrated the potential to discern audio features through the application of DNNs \cite{Frommholz23, comanducci2024}, and some are centered in the specific context of HRTF analysis and understanding of spatial audio \cite{Thuillier2018, zurale2022deep, zielinski2022, francl2022deep}. 

Building on this body of research, and particularly inspired by the approach in \cite{Thuillier2018}, the authors recently introduced a simple convolutional neural network (CNN) model designed to predict the elevation of a given HRTF pair \cite{ICA22SPAT}. The primary objective was to gain deeper insights into elevation perception by leveraging explainable artificial intelligence (XAI) techniques, including class activation mapping (CAM) \cite{Zhou2015} and gradient-based CAM (Grad-CAM) \cite{Selvaraju2017}. 
Although more complex DL models exist, we prioritized interpretability over complexity. Since our focus is on identifying key spectral patterns, the need for models that capture long sequences or complex relationships, such as transformers, is less critical for this task. Moreover, complex models often generate abstract and less transparent relationships between inputs and outputs, making it harder to directly trace the decision-making process. For example, in transformer architectures, which operate on embeddings, it can be difficult to explain predictions based on opaque, high-dimensional representations. 
Our goal is to understand primary elevation cues across a wide range of HRTF datasets by identifying key spectral features that improve spatial audio perception. To achieve this, we train a CNN model to classify HRTFs into different elevation sectors and use CAM to analyze the important features and frequency bands affecting prediction accuracy. This approach aims to uncover common characteristics and variations among subjects and measurement conditions, ultimately identifying consistent elevation cues across diverse HRTF datasets.

The paper is organized as follows: Section~\ref{sec:background} covers the background and problem formulation. Section~\ref{sec:datasets} provides an overview of the datasets and their measurement distinctions. Section~\ref{sec:methodology} details the methodology and experiments. Results and discussion are presented in Sections~\ref{sec:experiments} and \ref{sec:discussion}, respectively. The paper concludes with findings and future directions in Section~\ref{sec:conclusion}.

\section{Background}
\label{sec:background}

\subsection{Coordinate System and Elevation Classes}

\begin{figure}[t]
\begin{center}
	\includegraphics[width =0.8\columnwidth]{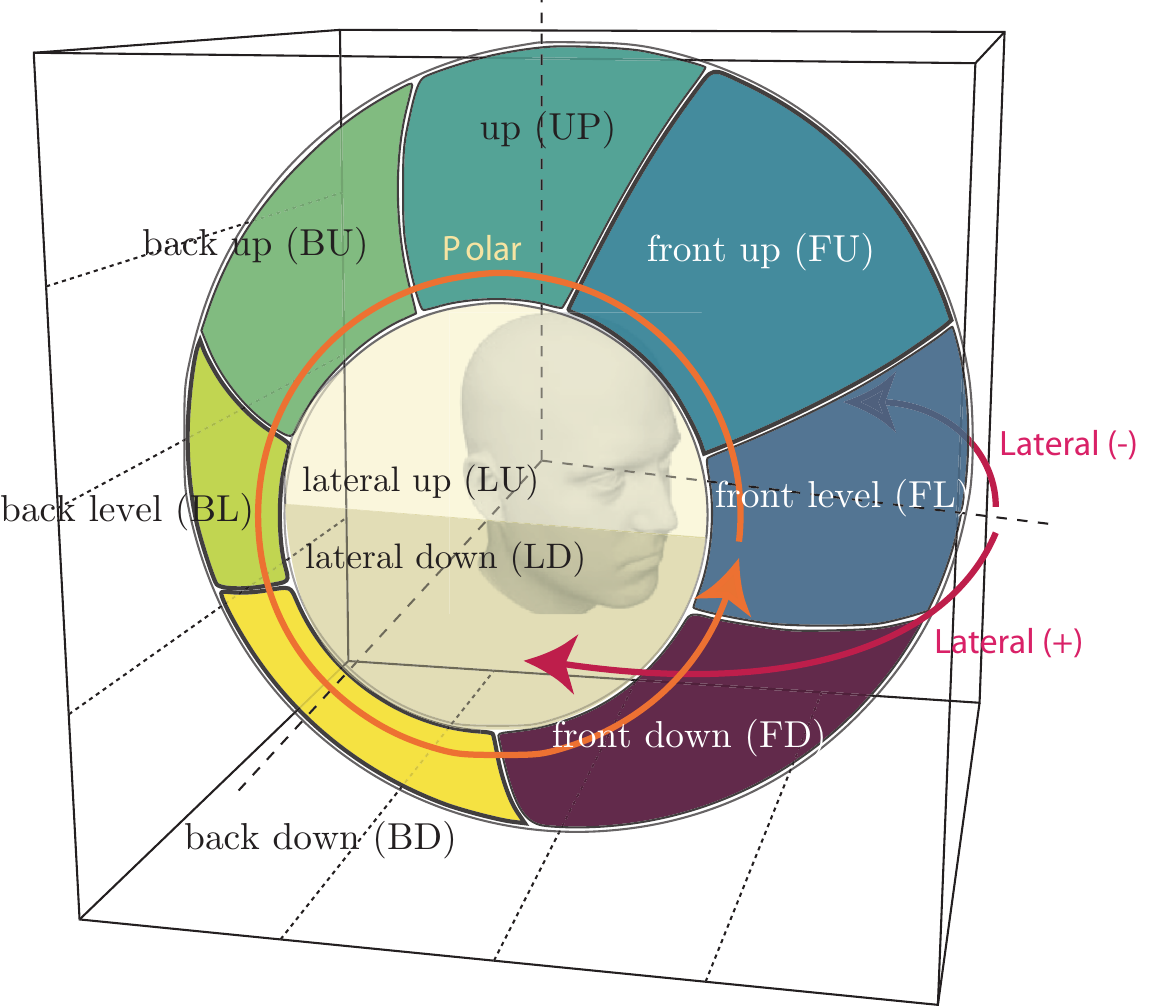}
	\caption{Horizontal-polar coordinate system and defined elevation sectors.}
 \label{fig:elevation_divisions}
	\end{center}
\end{figure}

To facilitate elevation classification, we adopted a horizontal-polar coordinate system, as shown in Figure~\ref{fig:elevation_divisions}. In this system, the horizontal position is determined by the lateral angle, which is the angle between the source, the listener's head center, and the vertical midline. The lateral angle spans from -90$^{\circ}$ (left) to 90$^{\circ}$ (right), with 0° at the vertical midline. The vertical and front/back position is defined by the polar angle, which represents the angle around the horizontal pole aligned with the listener's interaural axis. The polar angle ranges from 0$^{\circ}$ (on the front side), through 90$^{\circ}$ (above the interaural axis), through 180$^{\circ}$ (on the rear horizon), through 270$^{\circ}$ (beneath the interaural axis), to 360$^{\circ}$ (again on the front side). This coordinate system has a distinct advantage because it matches the two main dimensions influenced by the two types of localization cues. The lateral angle primarily deals with cues related to differences between the ears, while the polar angle combines the front/back and up/down aspects, which are shaped by the characteristics of the sound spectrum.

Similar to \cite{Thuillier2018}, our case study centers on categorizing a given HRTF pair into seven elevation classes, spanning from ``Front-Down" to ``Back-Down", as illustrated in Figure~\ref{fig:elevation_divisions}. The boundaries for each class are defined in Table~\ref{tab:ElevationSectors}.
Opting for a sector-based classification approach over regression enhances the interpretability and clarity of the model's decision-making process. Classification proves to be more resilient to noise and data variability, as categorizing elevation angles into sectors reduces the model's sensitivity to minor fluctuations in input features. This heightened robustness preserves stability in predictions, particularly when handling uncertainties, data inaccuracies, and the diversity of HRTFs from various datasets and acquisition conditions. While regression could better capture the continuous nature of spectral cues, it presents significant challenges in terms of interpretability with XAI techniques. Saliency maps for classification models are generally more reliable and easier to interpret, allowing for clearer insights into the model's behavior. Moreover, our experiments suggest that the spectral cues identified by the model remain consistent across different sector granularities, further supporting the validity and robustness of this approach.

\begin{table}[h!]
\caption{Categorization of angular directions into elevation classes}
\centering
\begin{tabular}{lcr}
\toprule
\textbf{Class} & \textbf{Polar Angle} & \textbf{Lateral Angle} \\
\midrule
Front Down & $(-90, -20]$ & $[-60, 60]$ \\
Front Level & $(-20, 20]$ & $[-60, 60]$ \\
Front Up & $(20, 70]$ & $[-60, 60]$ \\
Up & $(70, 110]$ & $[-60, 60]$ \\
Back Up & $(110, 160]$ & $[-60, 60]$ \\
Back Level & $(160, 200]$ & $[-60, 60]$ \\
Back Down & $(200, 270]$ & $[-60, 60]$ \\
Lateral Up & $[0, 180)$ & $|lateral| > 60$ \\
Lateral Down & $[-180, 0)$ & $|lateral| > 60$ \\
\bottomrule
\end{tabular}
\label{tab:ElevationSectors}
\end{table}

\subsection{CNNs  for classification}

In recent years, CNNs have surged in popularity, revolutionizing various fields with their ability to learn and extract meaningful patterns from data. This rise to prominence can be attributed to the extraordinary success of CNNs in tasks such as object detection \cite{O'Shea2015}, speech recognition \cite{Kwon2020}, acoustic scene classification \cite{ABEBER2020} or medical diagnostics \cite{Yadav2019}. What sets CNNs apart is their innate capability to automatically identify and capture relevant features within input data, enabling them to excel in tasks that involve complex patterns and hierarchical representations. In the context of elevation location from HRTF analysis, CNNs can analyze HRTF signals and capture important spectral cues and acoustic features associated with different elevation positions \cite{Thuillier2018}. The architecture of CNNs includes convolutional layers that employ filters to scan the HRTF input and capture local patterns convolved with learned weights. These filters act as feature detectors, enabling the network to identify crucial structures in HRTF data. By incorporating pooling layers, CNNs can reduce spatial dimensions while preserving essential features. Through training on labeled datasets, CNNs learn to extract discriminative features that contribute to accurate elevation classification. Moreover, it inherently integrates intraclass variability by generalizing across different examples within each elevation sector, focusing on the key features that remain consistent and relevant for classification despite observed variations.

\subsection{Class Activation Mapping}

When exploring the interpretability of CNNs, CAM emerges as a valuable technique for unraveling the decision-making process within these models \cite{Barredo2020}. By accentuating key regions in the input that influence the classification output of the model, CAM provides a way to study the saliency of individual components \cite{Zhou2015}. Utilizing feature maps from the final convolutional layer of the CNN, CAM conducts a weighted aggregation to pinpoint the most influential areas. In the specific context of HRTF analysis, the adaptation of CAM proves insightful in identifying specific regions or frequency bands crucial to the network's prediction of source elevation. These identified regions offer valuable insights into spatial or spectral cues influencing the decision of the model.

Mathematically, assuming a global average pooling operation and a dense prediction layer after the last convolutional layer of a CNN, the class-score for class $c$ before softmax normalization,  $Y^c \in \mathbb{R}$, can be written as:
\begin{equation}
    Y^c 
    = \sum_{k}w_{k}\left(\sum_{x,y}^{c}A_{k}(x,y)\right)
    = \sum_{k}\sum_{x,y}w_{k}^{c}A_{k}(x,y),
\end{equation}
where $w_k^{c}$ is the weight of the $k$-th channel unit for class $c$ in the last convolutional layer and $A_k(x,y)$ is its activation at location $(x,y)$. The class activation map for class $c$ is defined as:
\begin{equation}
    M^{c}(x,y) = \sum_{k}w_{k}^{c}A_{k}(x,y).
\end{equation}
Note that, since $Y^c=\sum_{x,y}M^c(x,y)$, $M^c(x,y)$ can be interpreted as the importance of the activation at $(x,y)$ that motivates classifying the input as a member of class $c$. By upsampling $M^c$ to the size of the input, the regions most relevant with regard to class $c$ can be identified in a simple way.

\begin{figure*}[t!]
\begin{center}
	\includegraphics[width =0.95\textwidth]{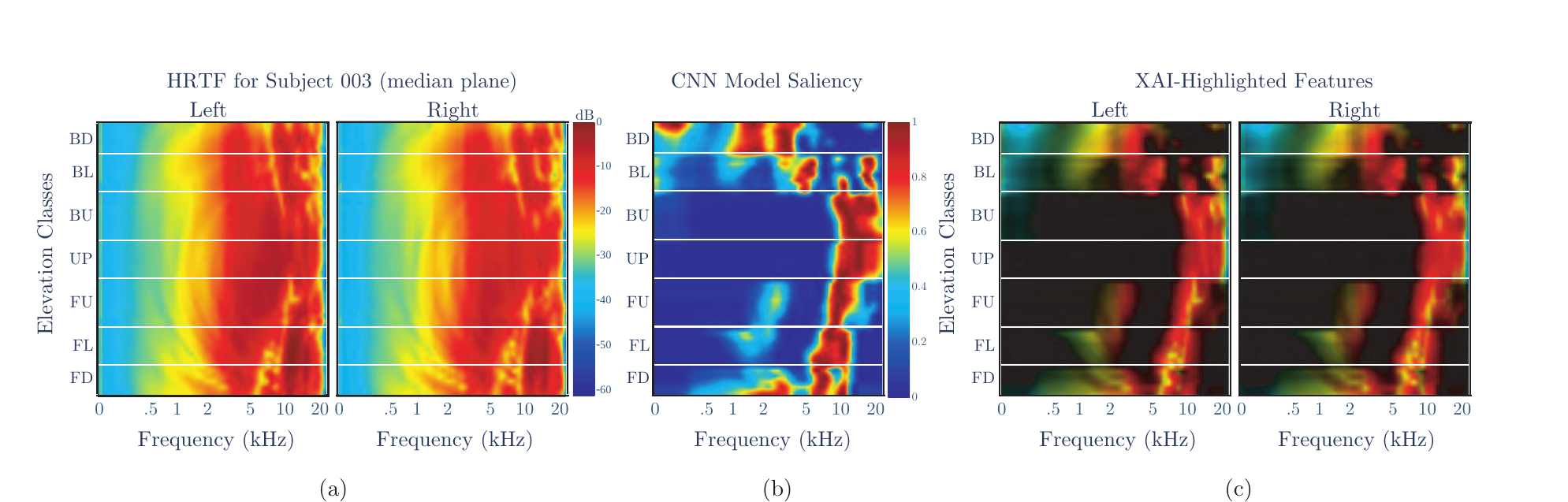}
	\caption{CAM saliency example on HRTFs from Subject 003 of the CIPIC dataset. (a) HRTF magnitude maps divided into the considered elevation sectors. (b) CAM-based saliency. (c) Saliency applied as occlusion mask to highlight relevant cues.}
 \label{fig:saliency}
	\end{center}
\end{figure*}

\subsection{Saliency}

The term \emph{saliency} is widely used as a measure of prominence or significance, often indicating the degree of attention or focus a particular element attracts within a given context. By applying CAM to the input of a CNN model, saliency can be obtained by exploring how $M^c$ distributes across the input of the model. High saliency within a sample indicates crucial cues for prediction, while low saliency suggests less influence on the model's decision. In Figure~\ref{fig:saliency}, we showcase a specific example of saliency from a CNN model trained on the CIPIC dataset. Specifically, (a) shows the HRTF maps (ipsilateral and contralateral) of subject 003 from such dataset. Each row in this map represents HRTF responses for a given lateral angle (0 in this case), ordered by polar angle along the y-axis. Horizontal lines group responses belonging to the same elevation class. Panel (b) displays the saliency map obtained after applying CAM, highlighting crucial regions for categorizing each HRTF into its corresponding elevation class. Notably, high saliency regions, often above 4 kHz, align with peaks and notches in the magnitude responses of the HRTFs. By applying the saliency map as an inverse opacity mask over the responses in (a), relevant features for each considered elevation sector are revealed and highlighted in (c). Note that the emphasized frequency regions vary across sectors, reflecting the model’s ability to prioritize different cues, which enhances its generalization across diverse HRTF datasets.

\subsection{Previous Work}

In our previous study \cite{ICA22SPAT}, CAM and Grad-CAM \cite{Selvaraju2017} were applied to a 1D-CNN trained for elevation classification, using the same elevation classes in Table~\ref{tab:ElevationSectors}. Results aligned with traditional research, highlighting a high-saliency region in the mid-to-high frequency range (4-10 kHz) \cite{IIDA2011, Alves-Pinto2014, Zonooz2019}. Additionally, saliency was observed in the low-frequency range (below 500 Hz) for backward regions (``back-down" to ``up"), but not for forward regions. Only the CIPIC dataset was used in this preliminary study. A contrasting saliency pattern was found between opposite regions, such as “front-level” and “back-level,” particularly in the 2-10 kHz range (compare FL/BL in Figure 7c). In a follow-up study \cite{ForumAccusticum23SPAT}, we explored HRTF pre-processing techniques to enhance inter-dataset model performance, but no universally effective pre-processing method was identified. However, significant performance improvements were observed when training on multiple datasets, even with limited samples.
In the following sections, we further investigate XAI techniques applied to a broad range of HRTF data, using models from previous works to uncover generalized HRTF cues that apply across diverse individuals. Aggregating data from multiple datasets helps mitigate the effects of outliers and prevents overfitting to non-generalizable features.

\section{Datasets}
\label{sec:datasets}
In this section, we provide a brief overview of the primary features of the 11 distinct HRTF datasets employed in this study, namely RIEC \cite{Watanabe2014}, FABIAN \cite{FABIAN}, CIPIC \cite{Algazi2001}, HUTUBS \cite{HUTUBS}, AACHEN \cite{aachen}, LISTEN \cite{LISTEN}, ARI \cite{ARI}, Crossmod \cite{crossmod}, SADIE \cite{SADIE}, BiLi \cite{BiLi} and 3D3A \cite{3D3A}.

\subsection{Recording Conditions}

A first analysis was carried out to identify key differences among the above datasets that could potentially hinder the compatibility of models trained on different data. To achieve this, we gathered information from various sources, including the datasets' websites, associated papers, and the information embedded inside the corresponding SOFA files. As anticipated, the analysis of these datasets unveiled a lack of standardization concerning both the average distance between the ears and the distance to the sound source. These non-uniformities can substantially influence the consistency of HRTF data due to the disparities arising from the varied interactions of sound waves with the listener. Additionally, we found that Swept Sines derived techniques were predominantly used for the source signal across most datasets, although alternative methods such as pseudo-random noise and the optimized Aoshima's time-stretched pulse (OATSP) \cite{OATSP} were also employed. It is important to note that the quality of the recorded HRTFs can be compromised by the presence of non-anechoic chambers, leading to undesired reflections (floor and walls) and distortions that can affect the accuracy and reliability of subsequent analysis or modeling based on these HRTFs. Table~\ref{tab:Recordings} summarizes the most important features of the datasets in terms of recording conditions.

\begin{table}[h!]
\caption{Differences in recording setup: ear-to-ear and source-to-listener distances (in meters), source signal measurement method and use of anechoic chamber.}
\small
\begin{threeparttable}
\begin{tabularx}{\linewidth}{lXXXX}
\toprule
\textbf{Dataset} & \textbf{Ear\tnote{*}} & \textbf{Source} & \textbf{Method\tnote{a}} & \textbf{Anechoic} \\ 
\midrule
3D3A & 0.085 & 0.76 & M.E.S.S. & Yes \\
AACHEN & 0.07 & 1.2 & Swept S. & Semi \\
ARI & 0.09 & 1.2 & Exp. S.S. & Semi \\
BiLi & 0.09 & 2.06 & Exp. S.S. & Yes \\
CIPIC & 0.09 & 1.5 & R. Noise & No\tnote{b} \\
Crossmod & 0.09 & 2.06 & Exp. S.S. & Yes \\
Fabian & 0.0662 & 1.7 & Swept S. & Yes \\
Hutubs & 0.076 & 1.47 & M.E.S.S. & Yes \\
Listen & 0.09 & 2.06 & Exp. S.S. & Yes \\
RIEC & 0.09 & 1.5 & OATSP & Yes \\
SADIE & 0.09 & 1.2 & Swept S. & Yes \\
\bottomrule
\end{tabularx}
\begin{tablenotes}
\item[*]Mean is shown, real individual measures declared in SOFA files or papers.
\item[a] OATSP: Optimized Aoshima’s Time-Stretched Pulse, Swept S: Swept Sine, R. Noise: Pseudo-Aleatory Noise, M.E.S.S.: Multiple Exponential Sine Sweep.
\item[b] Room with absorbers.
\end{tablenotes}
\end{threeparttable}
\label{tab:Recordings}
\end{table}

\subsection{Data Acquisition and Post-Processing}

The sampling rate (SR) and number of data samples per HRTF instance are also crucial considerations when integrating data from multiple sources for model training.  These variations play a significant role in the preprocessing stage, requiring careful attention to ensure consistency in the data representations used for training across diverse datasets. Table~\ref{tab:Samples} summarizes important information of the different datasets considered, including the number of subjects and the total number of HRTF responses. Moreover, certain datasets have undergone post-processing procedures prior to their publication. These post-processing techniques encompass various adjustments such as equalization, frequency cutoffs, numerical simulations, temporal windowing, gain calibration, low-frequency compensation, low-frequency extension, and diffuse field equalization. Table~\ref{tab:SampleProcessing} summarizes post-processing techniques reported in the considered datasets.

\begin{table}[h!]
\caption{Signal duration (in samples), sample rate and number of subjects across datasets.}
\begin{threeparttable}
\begin{tabularx}{\linewidth}{lXXXX}
\toprule
\textbf{Dataset} & \textbf{Duration} & \textbf{SR (kHz)} & \textbf{Subjects (used)} & \textbf{Total HRTFs} \\  
\midrule
3D3A & 2048 & 96 & 38 & 24624\\
AACHEN & 256 & 44.1 & 48 & 109440\\
ARI & 256 & 48 & 200 (97) & 150347\\
BiLi & 512 & 96 & 56 & 94080\\
CIPIC & 200 & 44.1 & 45 & 56250\\
Crossmod & 8192 & 44.1 & 24 & 15624\\
Fabian & 256 & 44.1 & 22 (11) & 131450\\
Hutubs & 256 & 44.1 & 96 & 42240\\
Listen & 8192 & 44.1 & 50 & 9350\\
RIEC & 512 & 48 & 105 & 90825\\
SADIE & 256 & 48 & 20 & 60584\\
\bottomrule
\end{tabularx}
\begin{tablenotes}
\item[*] Parentheses indicate numerically simulated responses.
\end{tablenotes}
\end{threeparttable}
\label{tab:Samples}
\end{table}

\begin{table}[h!]
\centering
\caption{Processing summary of datasets in SOFA files: Raw data, Equalization (Eq), Frequency Cut Off (FCO), Temporal Windowing (TW), Low Frequency Compensation (LFC), Gain Calibration (GC), Diffuse Field Equalization (DF Eq), Numerical Simulation (NS).}

\begin{tabular}{l|cccccccc}
\hline
\rotatebox[origin=C]{0}{\textbf{Dataset}} & \rotatebox[origin=C]{-90}{\textbf{Raw}} & \rotatebox[origin=C]{-90}{\textbf{Eq}} & \rotatebox[origin=C]{-90}{\textbf{FCO}} & \rotatebox[origin=C]{-90}{\textbf{TW}} & \rotatebox[origin=C]{-90}{\textbf{LFC}} & \rotatebox[origin=C]{-90}{\textbf{GC}} & \rotatebox[origin=C]{-90}{\textbf{DF Eq}} & \rotatebox[origin=C]{-90}{\textbf{NS}} \\ \hline
3D3A             &              & \cmark      &               &             &              &             &                 &             \\ \hline
AACHEN           &              &             & \cmark        &             &              &             &                 &             \\ \hline
ARI              &              &     \cmark        &        &  \cmark            &              &             &                 &             \\ \hline
BiLi             &              & \cmark      &               & \cmark      &              &             &                 &             \\ \hline
CIPIC            & \cmark       &             &               &             &              &             &                 &             \\ \hline
Crossmod         & \cmark       &             &               &             &              &             &                 &             \\ \hline
Fabian           &              &             &               &             &              &             &                 & \cmark      \\ \hline
Hutubs           &              &             & \cmark        &             &              &             &                 &             \\ \hline
Listen           & \cmark       &             &               &             &              &             &                 &             \\ \hline
RIEC             &              &             &               & \cmark      & \cmark       & \cmark      &                 &             \\ \hline
SADIE            &              &             &               & \cmark      & \cmark       &             & \cmark          &             \\ \hline
\end{tabular}
\label{tab:SampleProcessing}
\end{table}

\subsection{Spatial Distribution}

\begin{figure*}[h!]
    \centering
    \includegraphics[width=\linewidth]{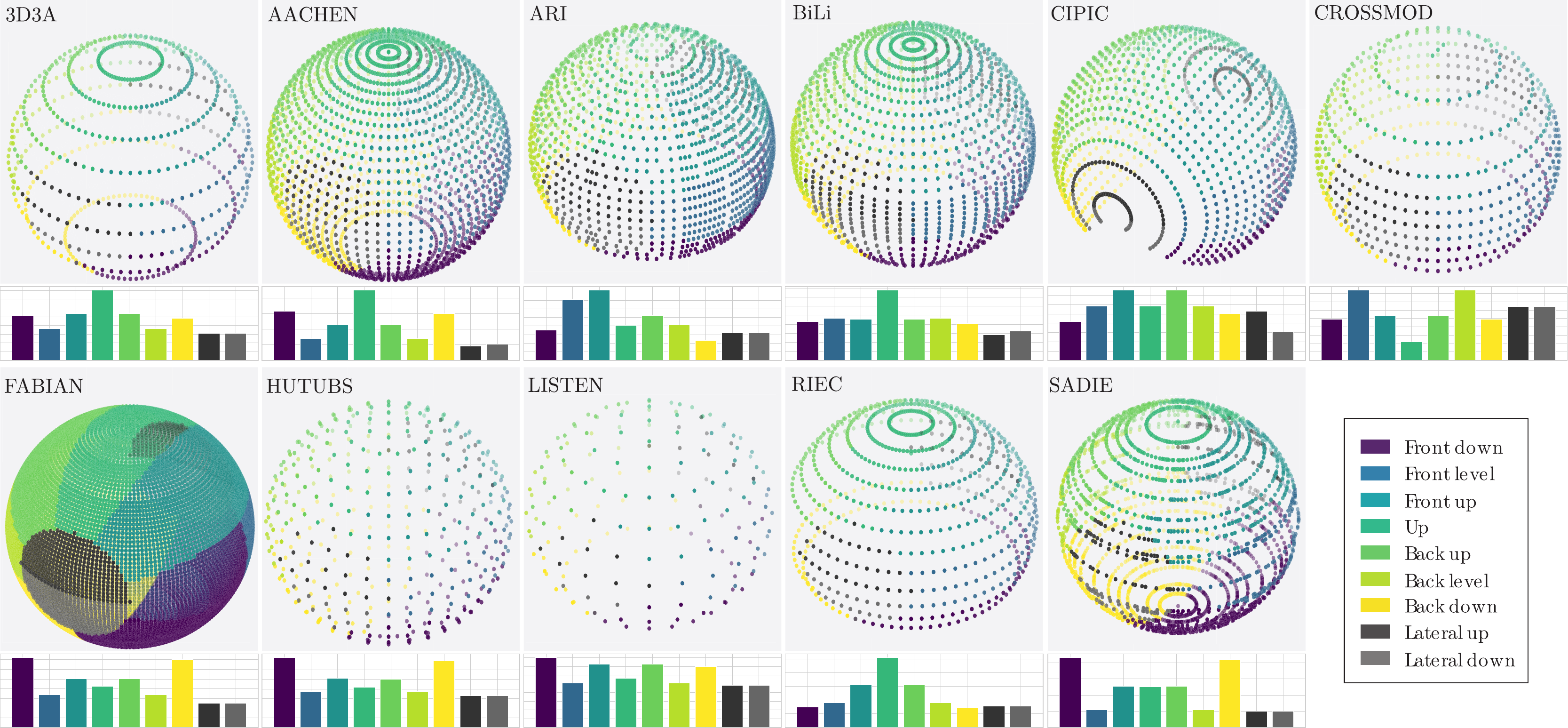}
    \caption{Spatial coordinates and class balance across datasets.}
    \label{fig:datasets}
\end{figure*}

The distribution of spatial measurements can exhibit significant variations among datasets. Discrepancies may arise in the density of samples, the range of elevation angles covered and the uniformity of sample distribution. Some datasets may display imbalances, with a higher concentration of samples along either the lateral or polar angles. Furthermore, variations in the class distribution of datasets can potentially influence the performance of HRTF models trained using such data. Figure~\ref{fig:datasets} shows graphically the spatial sampling and corresponding class balance across all the considered datasets.

\section{Methodology}
\label{sec:methodology}
 
\subsection{CNN Model}

\begin{table*}[ht]
\caption{Configuration of the CNN layers.}
\begin{center}
    \begin{tabular}{lrrrrrrrrrr}
    \toprule
    {}         & Input  &  Conv1 &  MPool1 &  Conv2 &  MPool2 &  Conv3 &  MPool3 &  Conv4 &  GAP &  Dense \\
    \midrule
    Out Dimension  & 257    &    257 &     128 &    128 &      64 &     64 &      32 &     32 &   16 &      9 \\
    Num Filters    &   2    &     64 &      64 &     32 &      32 &     32 &      32 &     16 &    1 &      1 \\
    Kernel Size &   0    &     16 &       2 &     16 &       2 &      8 &       2 &      8 &    0 &      0 \\
    Parameters &   0    &   2112 &       0 &  32800 &       0 &   8224 &       0 &   4112 &    0 &    153 \\
    \bottomrule
    \end{tabular}
\end{center}
\label{table:CNN_PARAMS}
\end{table*}

\begin{figure*}[h!]
    \centering
    \includegraphics[width=\linewidth]{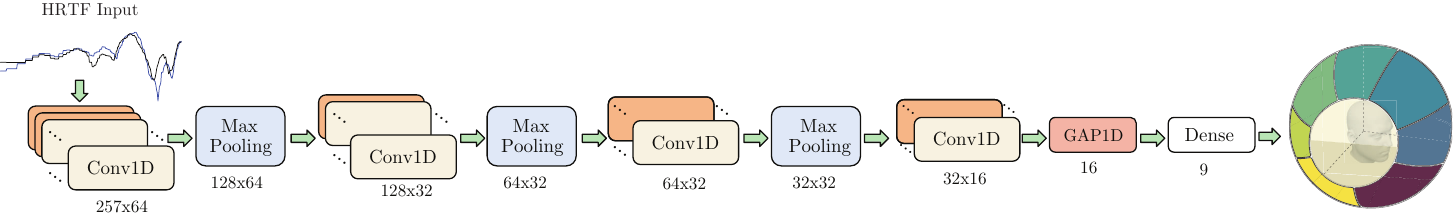}
    \caption{Architecture of the 1D-CNN model used for identification of spatial cues. Numbers indicate dimension (along the frequency axis) and number of filters, respectively.}
    \label{fig:model_cnn}
\end{figure*}

A fully convolutional model with a simple architecture is employed, as depicted in Figure~\ref{fig:model_cnn}. The design prioritizes achieving meaningful classification accuracy while maintaining simplicity to facilitate the application of common XAI techniques \cite{Ibrahim23, szczepankiewicz2023ground, amorim2023evaluating}. While deeper architectures might offer improved accuracy, they often reduce interpretability and increase the risk of overfitting, particularly when working with smaller datasets. Furthermore, given the challenge of validating saliencies without a well-defined ground truth in this domain, we prioritized interpretability over complexity to ensure the model delivers reasonable classification performance. The chosen model employs 1D convolutional layers to capture patterns across the frequency axis of HRTF data, which is central to our analysis. 

The model follows a conventional architecture, consisting of three 1D convolutional blocks with ReLU activation and max-pooling between blocks for frequency downsampling. 
This pooling strategy was chosen due to its effectiveness in preserving salient spectral features, as oposed to mean-pooling which might smooth out noise at the expense of a loss of sharpness in the detected features.
A final convolutional layer followed by global average pooling (GAP) is utilized to summarize the filter responses before the final dense layer, configured with softmax activation. Detailed information about the parameters corresponding to the different layers is shown in Table~\ref{table:CNN_PARAMS}. 

The model was trained using the Adam optimizer with an initial learning rate of $\eta = 10^{-4}$ with categorical cross-entropy as the loss function, with a reduction of the learning rate by a factor of 0.5 when the validation loss doesn't improve every 15 epochs. For the training process, a batch size of 32 and 200 epochs with early stopping were established to ensure stability and prevent overfitting. 
The data of each dataset was divided on a per-subject basis into training, validation, and test sets, ensuring that the data across partitions always belonged to different individuals. Approximately 80\% of the data was allocated for training, with 10\% reserved for validation and another 10\% for testing. Even though these unbalanced partitions may lead to high variance in performance estimates, especially for the smaller datasets, we prioritized having training sets that are as large and representative as possible, as our primary focus is on inter-dataset generalization and explainability.
We trained one individual model for each dataset plus a combined one with 10\% of the training data of each dataset for a total of 12 models.

\subsection{HRTF Pre-Processing}

To convert HRIR data from different datasets into HRTFs, we followed standard steps. Firstly, we resampled the original HRIR signals to a sampling rate of 44.1 kHz to account for variations in sampling rate across datasets. Then, we applied a Fast Fourier Transform (FFT) for frequency conversion, retaining only the magnitude information. In the context of classifying the elevation sector of a given input HRTF, the decision to prioritize the magnitude spectrum over phase information is grounded in the nature of the task. Elevation perception is primarily associated with the spatial characteristics embedded in the magnitude spectrum, which captures the amplitude or energy distribution at different frequencies \cite{Zonooz2019}. The HRTFs, comprising two channels (left and right), undergo further modification based on the spatial coordinates of the measurement. Specifically, we categorize them as ipsilateral (same side) and contralateral (opposite side) channels, swapping them as necessary. This adjustment ensures proper alignment with the spatial coordinate system used for the measurements. Additionally, in a previous work \cite{ForumAccusticum23SPAT} we conducted a comparison of different HRTF standardization techniques to enhance the performance of models when predicting elevation for samples from datasets other than those they were trained with. We tested several pre-processing alternatives working across four different aspects, which are outlined below.

\begin{itemize}
\item{\emph{Normalization}:}
we evaluated three normalization techniques: no normalization, min-max normalization, and average equator energy (AEE) normalization \cite{Zhang2022}. While min-max normalization was initially considered to standardize HRTF magnitudes, AEE normalization was ultimately chosen for the final experiments, as it yielded better results in all our preliminary experiments. This normalization involves dividing each HRTF sample by the average energy of HRTFs at the equator (zero elevation angle).

\item{\emph{Mel warping}:}
this technique involves transforming HRTFs from a linear scale to the Mel scale, aiming to provide a perception-oriented  frequency representation. Although the primary objective of this approach was not data standardization, we employed it to assess potential performance improvements in the trained models. To achieve this, we partitioned the frequency range into evenly spaced points on the Mel Scale and selected the frequency bins closest to their corresponding frequencies in Hz. This conversion enabled us to evaluate the model's performance in a frequency representation that aligns more closely with human auditory perception without changing the input size.

\item{\emph{Frequency cut-off}:}
we conducted tests using various effective frequency ranges, taking into account that certain datasets may have limited functional frequency ranges, and some may involve simulated or processed lower frequencies.

\item{\emph{HRTF amplitude scaling}:}
we carried out experiments using two different scales for HRTFs: linear and log10, where the log10 scale compresses the amplitude of the input signals to achieve a more perceptually uniform output.

\item{\emph{ERB filtering}:}
in this work, to gain deeper insights aligned with human perception, we tested an additional preprocessing technique based on Equidistant Rectangular Bandwidth (ERB) filtering, which mimics the cochlea's frequency response. Following the filter spacing and configuration from \cite{francl2022deep}, we implemented a filterbank consisting of 255 filters with center frequencies ranging from 50 Hz to 22,050 Hz.
\end{itemize}

\subsection{Pre-processing Configurations}
\label{sec:preprocessing}

Although our experiments explored numerous combinations of the aforementioned HRTF pre-processing techniques, we selected the following key configurations for discussion:


\begin{itemize}
\item[a)]\emph{No Pre-processing (Raw)}: a baseline consisting of raw HRTF data without further standardization or processing techniques, as provided in the original datasets.


\item[b)] \emph{Optimized Pre-processing}:
while no single pre-processing technique consistently improved performance across all datasets, certain combinations performed better than others. This particular configuration was selected for demonstrating the best overall performance compared to all other tested configurations, considering both classification accuracy and generalization across datasets.

\item[c)] \emph{Perceptually-based Pre-processing (Perceptual)}:
since this work prioritizes the explainability of elevation cues over classification accuracy alone, we identified one of the best-performing configurations among those incorporating a perceptual-based representation. This configuration, based on ERB filtering, was selected for its balance between competitive performance and enhanced interpretability from a human auditory perspective, despite a slight trade-off in accuracy.

\end{itemize}

The pre-processing settings for each of the above configurations are summarized in Table~\ref{tab:SetsParameters}. For each set of parameters, we calculated the HRTFs and trained individual models for each dataset using its corresponding training set of samples. The subsequent section will provide a detailed account of the comprehensive testing conducted within each pre-processing configuration.

\begin{table}[h!]
\caption{Summary of Preprocessing Configurations }
{\small
\begin{tabularx}{\columnwidth}{l|lllll}
\toprule
\textbf{Config} & \textbf{Norm.} &  \textbf{\makecell[l]{Amplitude\\ Scale}}  & \textbf{\makecell[l]{Bands\\Cut}} & \textbf{\makecell[l]{Freq.\\axis}} \\
\midrule
\bf Raw & \xmark & Linear & None & Linear \\
\bf Perceptual & \makecell[l]{AEE}  & Linear & \makecell[l]{50Hz$\sim$22 kHz} & ERB\\
\bf Optimized & \makecell[l]{AEE}  & Linear & \makecell[l]{50Hz$\sim$22 kHz} & Linear \\

\bottomrule
\end{tabularx}
}%
\label{tab:SetsParameters}
\end{table}

\section{Experiments}
\label{sec:experiments}

\begin{figure}[h!]
    \centering
    \includegraphics[width=0.95\columnwidth]{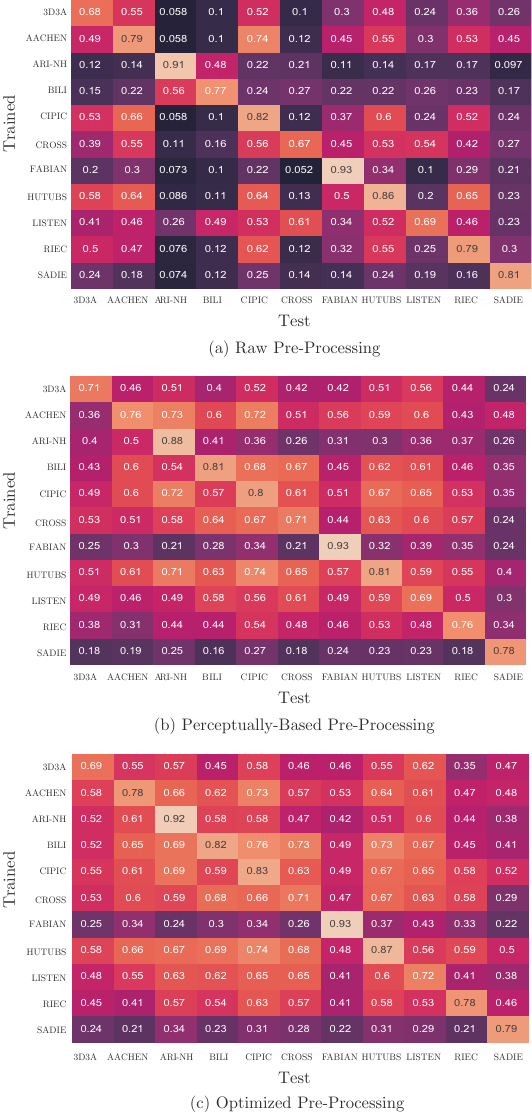}
    \caption{F1-Score of models trained on a specific dataset (rows) when tested over a different dataset (columns).
    }
    \label{fig:Matrix_Confussion_Accuracies}
\end{figure}


\newcommand{\taulaEnorme}{%

\begin{table*}[h!]
\caption{\textcolor{red}{Mean Accuracy F1 Scores. Column Ref. Train shows the performance of a model trained on the Ref. Dataset against all datasets. Column Ref. Test shows the performance of all models against Ref. Dataset}}
\centering
\begin{tabularx}{\textwidth}{l|XX|XX|XX|}
\hline
\textbf{Ref. Dataset} & \multicolumn{2}{c|}{\textbf{RAW}} & \multicolumn{2}{c|}{\textbf{Optimized Pre-processing}} & \multicolumn{2}{c|}{\textbf{Perceptual Based}} \\
\hline
 & \textbf{Ref. Train} & \textbf{Ref. Test} & \textbf{Ref. Train} & \textbf{Ref. Test} & \textbf{Ref. Train} & \textbf{Ref. Test} \\
\hline

\textbf{3D3A}                             & 0.332                  $\pm$ 0.206                 & 0.392                  $\pm$ 0.187                & 0.522                  $\pm$ 0.094                 & 0.490                 $\pm$ 0.135                 & 0.471                  $\pm$ 0.117                 & 0.430                  $\pm$ 0.143                \\
\textbf{AACHEN}                           & 0.415                  $\pm$ 0.246                 & 0.450                  $\pm$ 0.215                & 0.606                  $\pm$ 0.095                 & 0.543                 $\pm$ 0.163                 & 0.575                  $\pm$ 0.127                 & 0.481                  $\pm$ 0.165                \\
\textbf{ARI-NH}                           & 0.251                  $\pm$ 0.241                 & 0.211                  $\pm$ 0.276                & 0.548                  $\pm$ 0.145                 & 0.598                 $\pm$ 0.181                 & 0.400                  $\pm$ 0.174                 & 0.551                  $\pm$ 0.205                \\
\textbf{BiLi}                             & 0.303                  $\pm$ 0.188                 & 0.242                  $\pm$ 0.229                & 0.628                  $\pm$ 0.138                 & 0.556                 $\pm$ 0.172                 & 0.565                  $\pm$ 0.135                 & 0.502                  $\pm$ 0.183                \\
\textbf{CIPIC}                            & 0.387                  $\pm$ 0.256                 & 0.488                  $\pm$ 0.220                & 0.619                  $\pm$ 0.094                 & 0.618                 $\pm$ 0.164                 & 0.592                  $\pm$ 0.123                 & 0.564                  $\pm$ 0.178                \\
\textbf{Crossmod}                         & 0.424                  $\pm$ 0.177                 & 0.231                  $\pm$ 0.210                & 0.582                  $\pm$ 0.118                 & 0.545                 $\pm$ 0.160                 & 0.556                  $\pm$ 0.130                 & 0.483                  $\pm$ 0.194                \\
\textbf{Fabian}                           & 0.256                  $\pm$ 0.243                 & 0.376                  $\pm$ 0.221                & 0.366                  $\pm$ 0.196                 & 0.483                 $\pm$ 0.169                 & 0.347                  $\pm$ 0.202                 & 0.490                  $\pm$ 0.177                \\
\textbf{HUTUBS}                           & 0.419                  $\pm$ 0.274                 & 0.459                  $\pm$ 0.205                & 0.638                  $\pm$ 0.110                 & 0.590                 $\pm$ 0.156                 & 0.616                  $\pm$ 0.115                 & 0.528                  $\pm$ 0.177                \\
\textbf{LISTEN}                           & 0.455                  $\pm$ 0.139                 & 0.289                  $\pm$ 0.173                & 0.554                  $\pm$ 0.116                 & 0.574                 $\pm$ 0.121                 & 0.523                  $\pm$ 0.100                 & 0.522                  $\pm$ 0.141                \\
\textbf{RIEC}                             & 0.374                  $\pm$ 0.231                 & 0.417                  $\pm$ 0.200                & 0.539                  $\pm$ 0.107                 & 0.473                 $\pm$ 0.154                 & 0.469                  $\pm$ 0.120                 & 0.466                  $\pm$ 0.147                \\
\textbf{SADIE}                            & 0.233                  $\pm$ 0.200                 & 0.296                  $\pm$ 0.191                & 0.312                  $\pm$ 0.166                 & 0.445                 $\pm$ 0.146                 & 0.263                  $\pm$ 0.175                 & 0.360                  $\pm$ 0.159                \\

\hline
\textbf{MEAN}                             & 0.350                  $\pm$ 0.218                 & 0.350                  $\pm$ 0.212                & 0.538                  $\pm$ 0.125                 & 0.538                 $\pm$ 0.156                 & 0.489                  $\pm$ 0.138                 & 0.489                  $\pm$ 0.170  \\             
\hline
\end{tabularx}

\label{tab:interdataset_performance}
\end{table*}
}

\newcommand{\AltraTaulaEnorme}{%
\begin{table*}[h!]
\caption{\textcolor{red}{Macro-Averaged Performance Metrics by Class and Pre-processing configuration. (Combined Model against each dataset on test partition). The best results for each row are highlighted in bold.}}
\centering
\begin{tabularx}{\textwidth}{l*{9}{>{\centering\arraybackslash}X}}
\toprule
\textbf{Class} & \multicolumn{3}{c}{\textbf{No Pre-processing}} & \multicolumn{3}{c}{\textbf{\textcolor{blue}{Optimized Pre-processing}}} & \multicolumn{3}{c}{\textbf{\textcolor{blue}{Perceptual Based}}} \\
\cmidrule(lr){2-4} \cmidrule(lr){5-7} \cmidrule(lr){8-10}
 & \textbf{Precision} & \textbf{Recall} & \textbf{F1} & \textbf{Precision} & \textbf{Recall} & \textbf{F1} & \textbf{Precision} & \textbf{Recall} & \textbf{F1} \\

\midrule

Mean & 0.670 $\pm$ 0.083 & 0.659 $\pm$ 0.092 & 0.655 $\pm$ 0.087      &\bf 0.715 $\pm$ 0.093 &\bf 0.711 $\pm$ 0.102 &\bf 0.706 $\pm$ 0.099       &   0.647 $\pm$ 0.097  &  0.643 $\pm$ 0.105  &  0.637 $\pm$ 0.102 \\
\midrule

Front Down & 0.751 $\pm$ 0.104 & 0.719 $\pm$ 0.079 & 0.729 $\pm$ 0.059       &\bf 0.787 $\pm$ 0.099 &\bf 0.740 $\pm$ 0.091 &\bf 0.759 $\pm$ 0.078        &  0.733 $\pm$ 0.113  & 0.702  $\pm$ 0.063  &  0.714 $\pm$ 0.076\\
Front Level & 0.664 $\pm$ 0.122 & 0.729 $\pm$ 0.143 & 0.688 $\pm$ 0.111       &\bf 0.754 $\pm$ 0.093 &\bf 0.761 $\pm$ 0.132 &\bf 0.753 $\pm$ 0.105        &  0.654 $\pm$ 0.133  &  0.683 $\pm$ 0.140  &  0.664 $\pm$ 0.128\\
Front Up & 0.663 $\pm$ 0.109 & 0.713 $\pm$ 0.119 & 0.682 $\pm$ 0.095       &\bf 0.723 $\pm$ 0.104 &\bf 0.782 $\pm$ 0.118 &\bf 0.746 $\pm$ 0.090        &  0.661 $\pm$ 0.114  &  0.696 $\pm$ 0.168  &  0.664 $\pm$ 0.122 \\
Up & 0.670 $\pm$ 0.113 & 0.654 $\pm$ 0.142 & 0.651 $\pm$ 0.091          &\bf 0.737 $\pm$ 0.120 &\bf 0.675 $\pm$ 0.099 &\bf 0.699 $\pm$ 0.083        &  0.680 $\pm$ 0.121  &  0.565 $\pm$ 0.113  &  0.607 $\pm$ 0.084 \\
Back Up & 0.695 $\pm$ 0.105 & 0.686 $\pm$ 0.101 & 0.685 $\pm$ 0.088          &\bf 0.727 $\pm$ 0.109 &\bf 0.718 $\pm$ 0.114 &\bf 0.719 $\pm$ 0.100         & 0.644 $\pm$ 0.089  &  0.696 $\pm$ 0.113  &  0.665 $\pm$ 0.085\\
Back Level & 0.690 $\pm$ 0.104 & 0.574 $\pm$ 0.149 & 0.619 $\pm$ 0.123        &\bf 0.717 $\pm$ 0.120 &\bf 0.632 $\pm$ 0.172 &\bf 0.665 $\pm$ 0.143        &  0.647 $\pm$ 0.119  &  0.585 $\pm$ 0.146  &  0.607 $\pm$ 0.116\\
Back Down & 0.714 $\pm$ 0.105 & 0.677 $\pm$ 0.115 & 0.684 $\pm$ 0.069         &\bf 0.743 $\pm$ 0.095 &\bf 0.788 $\pm$ 0.109 &\bf 0.762 $\pm$ 0.089        &  0.699 $\pm$ 0.107  &  0.711 $\pm$ 0.115  &  0.699 $\pm$ 0.088\\
Lateral Up & 0.592 $\pm$ 0.173 & 0.595 $\pm$ 0.186 & 0.581 $\pm$ 0.155         &\bf 0.622 $\pm$ 0.160 &\bf 0.675 $\pm$ 0.167 &\bf 0.634 $\pm$ 0.139        &   0.555 $\pm$ 0.197 & 0.555 $\pm$ 0.228 & 0.544 $\pm$ 0.209 \\
Lateral Down & 0.596 $\pm$ 0.131 & 0.588 $\pm$ 0.172 & 0.581 $\pm$ 0.132     &\bf 0.628 $\pm$ 0.175 &\bf 0.626 $\pm$ 0.158 &\bf 0.620 $\pm$ 0.156        &   0.551 $\pm$ 0.176 & 0.593 $\pm$ 0.163 & 0.567 $\pm$ 0.166\\

\midrule
\makecell[l]{Mean against\\unseen dataset\\(SONICOM \cite{SONICOM2023Engel})} & 0.512 & 0.490 & 0.482        & 0.526 &\ 0.536 & 0.512       &\bf   0.544  &\bf  0.542  &\bf  0.526 \\
\bottomrule
\end{tabularx}
\label{tab:PerformanceMetrics}
\end{table*}
}

\newcommand{\taulaSonicomVella}{%
\begin{tabularx}{\columnwidth}{l|X|X}
\toprule
\textbf{Class} & \textbf{\textcolor{blue}{Optimized Pre-processing}} & \textbf{\textcolor{blue}{Perceptual Based}} \\
\midrule
Mean & \bf 0.706 &  0.637 \\
\midrule
\makecell[l]{Mean against\\unseen dataset\\(SONICOM \cite{SONICOM2023Engel})} & 0.512 & \bf 0.526 \\
\bottomrule
\end{tabularx}
}

\newcommand{\taulaDatasets}{%
\begin{tabularx}{\columnwidth}{ll|ccc}
\toprule
\textbf{}                && \textbf{~Raw~} & \textbf{Perceptual} & \textbf{Optimized} \\ 
\midrule
\multicolumn{2}{l|}{In-domain} & 0.79        & 0.79                   & \textbf{0.80}                                           \\ 
\midrule
\multirow{4}{*}{\makecell[l]{Cross-domain}}
&\footnotesize best                                    & 0.54                              & 0.58                                           & \textbf{0.62} \\
&\footnotesize median &                        0.31                              & 0.47                                           & \textbf{0.53}                                          \\
&\footnotesize average &                       0.31                              & 0.46                                           & \textbf{0.51}                                          \\
&\footnotesize worst                                     & 0.10                              & 0.29                                           & \textbf{0.37}     \\
\bottomrule
\end{tabularx}
}

\newcommand{\taulaCombined}{%
\begin{tabularx}{\columnwidth}{ll|ccc}
\toprule
\textbf{} &
& \textbf{~~~Raw~~~}
& \textbf{Perceptual}
& \textbf{Optimized} 
\\
\midrule
%
\multirow{4}{*}{\makecell[l]{In-domain}}
& \footnotesize best    & 0.77 
&  0.77 
& \bf 0.80 
\\ & \footnotesize median  & 0.67 
&  0.70 
& \bf 0.75 
\\ & \footnotesize average & 0.66 
&  0.64 
& \bf 0.71 
\\ & \footnotesize worst   & 0.49 
&  0.47 
& \bf 0.51 
\\ \midrule
\multicolumn{2}{l|}{\makecell[l]{Cross-domain}} & 0.48 
& \bf 0.53 
& 0.51 
\\ \bottomrule
\end{tabularx}

}

\subsection{Average Inter-Dataset Performance}

\begin{table}[t]
\caption{
Average F1 performance  across models trained on all datasets, evaluated within the same domain (first row) and across different domains/datasets (next four rows).}
\taulaDatasets
\label{tab:summary_interdataset_performance}
\end{table}

Figure~\ref{fig:Matrix_Confussion_Accuracies} presents the normalized dataset-wise confusion matrices for each pre-processing configuration described in Section~\ref{sec:preprocessing}. Each matrix illustrates the average model performance (macro-averaged F1-Score) when trained on a specific dataset and tested on the test sets of other datasets. The diagonal represents the model’s performance when trained and tested on the same domain (i.e. dataset), while the off-diagonal entries assess the the robustness of features learned from one dataset when applied to others. This evaluation is crucial for spatial audio applications, where models must perform reliably beyond a single dataset or domain. Overall, models achieve reasonable performance within their native dataset but struggle to maintain performance under varying degrees of domain mismatch.


The results from Figure~\ref{fig:Matrix_Confussion_Accuracies} are summarized in Table~\ref{tab:summary_interdataset_performance}, where performance values are averaged across all models and datasets. The first row reports the average of the diagonal elements, reflecting performance when training and testing on independent subsets of the same dataset—representing models used in a single, relatively homogeneous domain. The second to fifth rows provide the maximum, median, average, and minimum of the off-diagonal elements in each row, then averaged across all models.


Both Figure~\ref{fig:Matrix_Confussion_Accuracies} and Table~\ref{tab:summary_interdataset_performance} highlight that while pre-processing does not significantly impact a model’s performance on its native dataset, it greatly influences its ability to generalize to unseen datasets. Notably, the optimized configuration outperforms the other two, though the perceptually-based one exhibits a similar trend. Additionally, both optimized and perceptual models largely agree on the best-performing datasets (CIPIC and Hutubs) and the most challenging ones (Fabian and SADIE), which also result in the weakest models when used for training (evidenced by the darkest rows in the confusion matrices).




\begin{table}[h!]
\caption{F1 Score of the Combined model (trained on all datasets), evaluated on individual datasets (first four rows) and on an unseen domain/dataset\protect\footnotemark{} (last row), for all pre-processing configurations (columns).}
\centering
\taulaCombined
\label{tab:summary_PerformanceMetrics}
\end{table}

\footnotetext{The Combined model has been tested on the unseen SONICOM dataset~\cite{SONICOM2023Engel}.}

\subsection{Combined Model}



In the previous section, we demonstrated that models trained on narrow domains exhibit varying performance when applied to data from different domains. This suggests that each model learns internal representations tailored to the specific characteristics of its training dataset. To develop a more generalizable model capable of learning broader features, we created a combined training set by sampling 10\% from each of the 11 datasets. The resulting model, trained under the same conditions as before across all configurations, is referred to as the \emph{Combined model}. This model was evaluated on the test partitions of all 11 datasets, meaning the training and test data come from the same domain in this case.


To further assess generalization and the impact of different pre-processing strategies, we conducted an additional evaluation on an entirely unseen dataset, SONICOM \cite{SONICOM2023Engel}, which includes data from over 200 subjects. This analysis tests the model’s robustness when faced with previously unseen data. The performance results are summarized in Table~\ref{tab:summary_PerformanceMetrics}. The first four rows reflect performance when both training and test data come from the same domain. Notably, even more clearly than before, the models with Perceptual and Optimized configurations agree on the worst-performing test dataset (Fabian) and the median (Crossmod). Although the Optimized configuration achieves the highest performance on Hutubs, CIPIC —favored by the Perceptual model— follows closely as the second best
with a F1-score of 0.79.
Overall, the Combined model exhibits a performance pattern similar to that of the individual models in their respective domains.

However, as shown in the last row of Table~\ref{tab:summary_PerformanceMetrics}, when tested on entirely new data, performance drops again to similar levels as before. Interestingly, the perceptually-based configuration using ERB filtering performs slightly better on the unseen dataset, suggesting that its learned internal representations are more aligned with generalizable auditory features.

\begin{figure}[t!]
    \centering
    \includegraphics[width=\columnwidth]{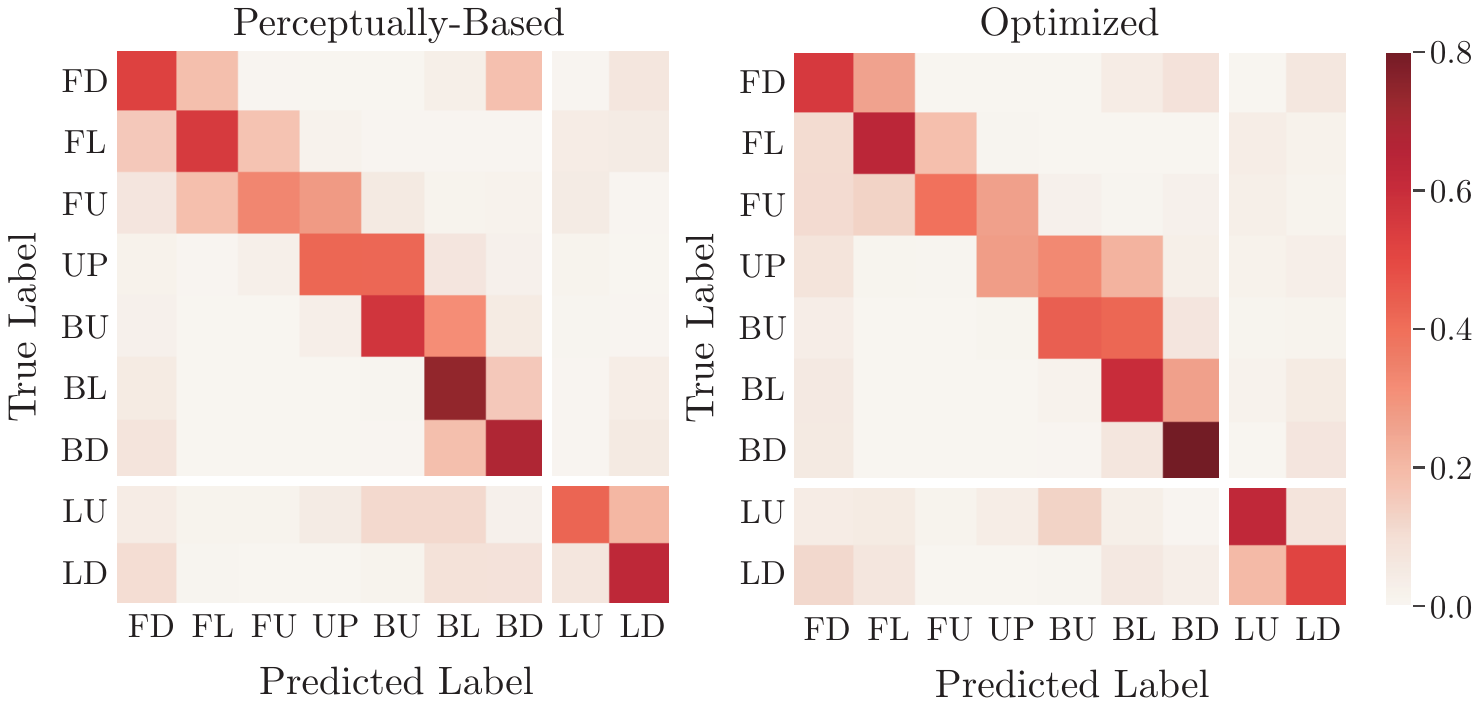}
    \caption{Confusion matrices showing class-specific accuracies and errors for Perceptual and Optimized models against SONICOM dataset.
    }
    \label{fig:CONF}
\end{figure}

In terms of class-wise performance, the metrics for most classes are quite similar, with the exception of the UP class, which performs somewhat worse. It is important to highlight that the Combined model, across both configurations, predominantly predicts either the correct orientation or one that is closely adjacent. This pattern, demonstrated in Figure~\ref{fig:CONF} for the SONICOM dataset, also reflects the overall class-wise performance trends observed in the models.


\subsection{Cross-Dataset Saliency}

In this section, we analyze the saliencies produced by the Combined model when trained with both the optimized pre-processing configuration and the perceptually-based configuration. For clarity, we present saliency results for three representative datasets, showcasing the worst (Fabian), median (Crossmod), and (consensual) best (CIPIC) cases. The goal is to identify common saliency patterns across subjects from different datasets that may be associated with underlying elevation cues.

\begin{figure*}[!t]
    \centering
    \includegraphics[width=\linewidth]{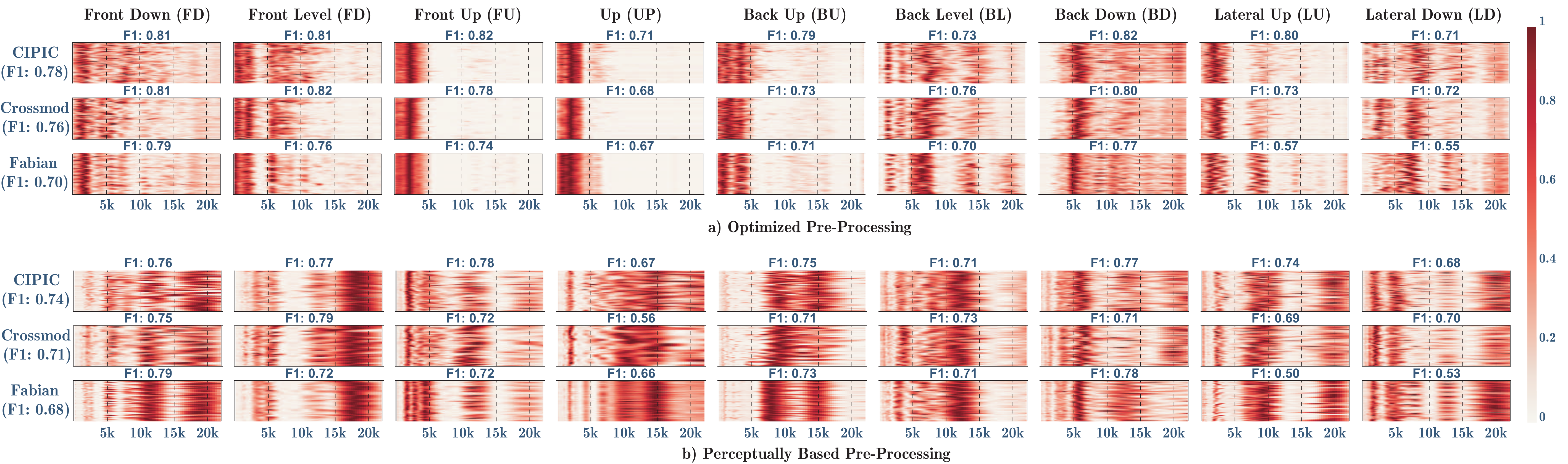}
    \caption{Aggregated saliency countour maps per class of the Combined model against datasets representative for Worst, Median and Best performance results. a) Using Optimized Pre-Processing. b) Using Perceptual Based Preprocessing }
    \label{fig:SALIENCY_BANDS_COMBINED_MODEL_AGAINST_ALL}
\end{figure*}

Figure~\ref{fig:SALIENCY_BANDS_COMBINED_MODEL_AGAINST_ALL} shows a matrix of panels, where the panel at each cell contains an image conformed by aggregating the saliencies resulting for all the HRTF responses corresponding to a given class (cell column) pertaining to a given dataset (cell row) that were correctly classified. The images are heatmaps where the vertical axis represents different HRTF responses (from all the subjects in the dataset) while the horizontal axis is frequency. The colormap assigns a darker color to higher saliency values, making visible which frequencies have a higher weight in the model's decision across individual responses for a given elevation class and as a function of its origin dataset. Additionally, the F1-score obtained for each elevation class at each dataset is provided a the top of each panel, and the overall F1-score for each dataset is shown below its name. 

Remarkably, saliencies associated with a specific class show strong alignment within each individual pre-processing configuration, indicating a meaningful consensus across datasets. This consistency highlights the model’s ability to reliably identify key frequency ranges essential for classifying responses into distinct elevation sectors. However, significant differences emerge when comparing the saliency maps of the two pre-processing configurations in Figure~\ref{fig:SALIENCY_BANDS_COMBINED_MODEL_AGAINST_ALL}a and Figure~\ref{fig:SALIENCY_BANDS_COMBINED_MODEL_AGAINST_ALL}b. While variations within a single preprocessing approach—such as additional salient zones or low-saliency regions within a shared frequency band—are present across datasets, the discrepancies between the two preprocessing methods are far more pronounced.

We hypothesize that while these dataset-specific frequency bands may reflect spectral cues arising from unique characteristics of certain datasets—potentially influenced by special measurement conditions—the choice of input representation plays a crucial role in shaping the model's decision-making process.Given the observed discrepancies, next sections will focus on the perceptually-based pre-processing, as it better aligns with human auditory perception. Unlike the optimized pre-processing, which tends to concentrate saliency in low-frequency regions, the perceptually-based approach distributes saliency more evenly, with higher weights at higher frequencies—aligning with widely accepted theories of elevation cue encoding. This choice enhances interpretability and improves the explainability of the model’s decisions.

\subsection{Prototype Saliency Contours}
The primary aim of this section is to generate generalized saliency contour prototypes for each elevation class that effectively capture the prevailing trends observed across all the considered datasets.These prototypes can be seen as spectral indicators, highlighting the relative significance of features along the frequency axis within their corresponding elevation sectors. 
%


As the set of saliencies corresponding to well classified samples according to the Combined Model (with perceptually-based pre-processing) and shown in Figure~\ref{fig:SALIENCY_BANDS_COMBINED_MODEL_AGAINST_ALL}, 
has different sizes for different datasets, they have been downsized to the minimum size by discarding saliencies that correspond to the lowest confidence levels. In this way, the same amount of information from each database is used in the following steps.

\begin{figure*}[h!]
    \centering
    
    \includegraphics[width=\textwidth]{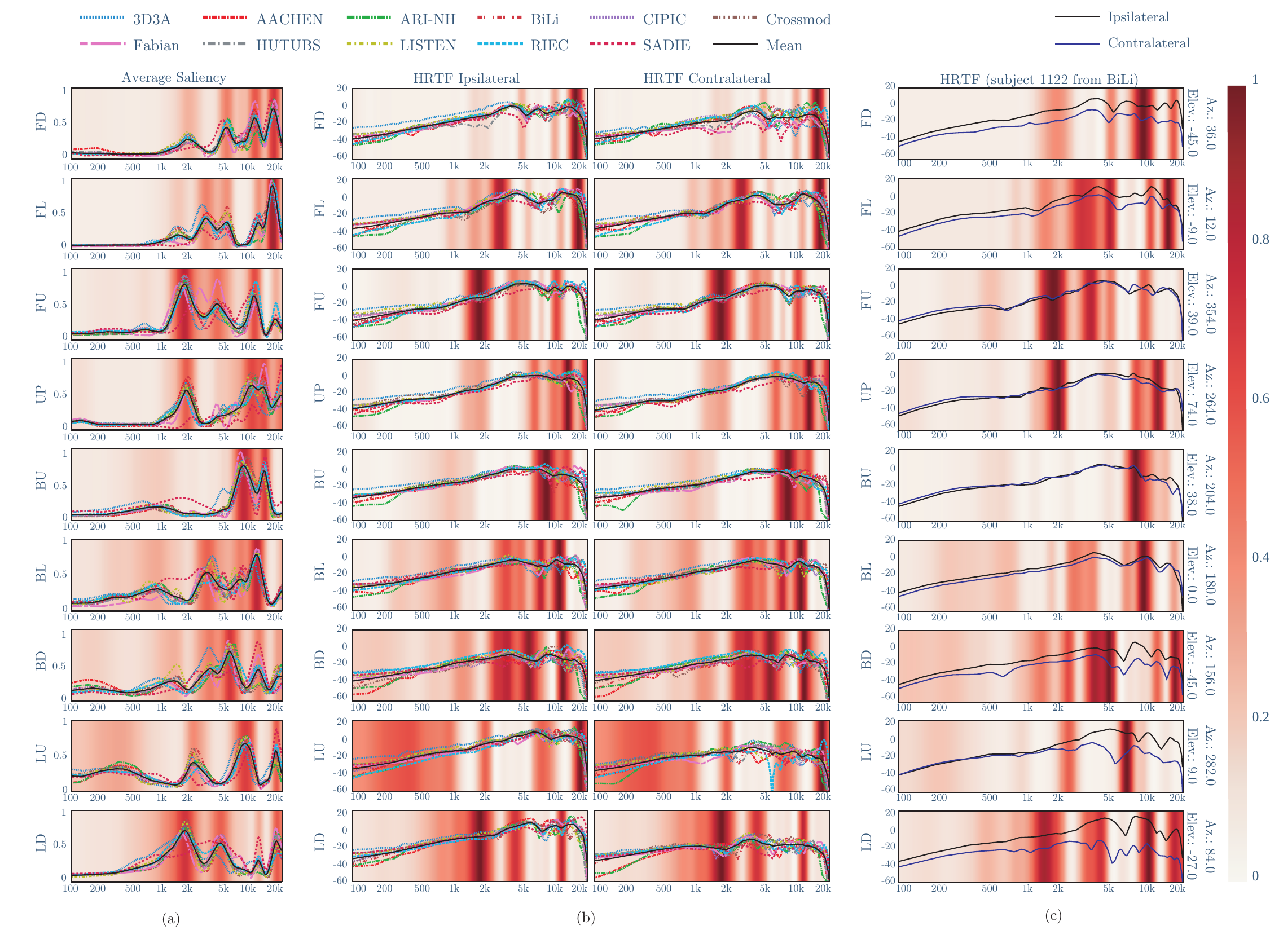}

    \caption{Saliency of Combined model (depicted always as a background heatmap). (a) Class-conditional averaged saliency contours for different datasets (line colors) and MSC (solid black line and background heatmap). (b) HRTF and corresponding saliency in the background for each elevation class. Line colors depict closest samples from each dataset.  (c) HRTF of the most representative subject (BiLi 1122) and corresponding saliencies as background.}
    \label{fig:MEAN_SALIENCY_BANDS_CombinedModel_AGAINST_EACH_DATASET}
\end{figure*}

Figure~\ref{fig:MEAN_SALIENCY_BANDS_CombinedModel_AGAINST_EACH_DATASET}a) shows the average saliency per class 
where different line colors indicate different datasets. 
The mean saliency contour (MSC), averaged across all datasets is shown both as a black solid line and as a colored background that varies along the frequency axis only.
We observe consistent saliency patterns across datasets, as evidenced by the substantial alignment of these average saliencies from different datasets.

To gain more insight about how MSC relates to particular saliencies from specific inputs, idealized representative HRTFs for each class have been constructed by averaging the ones corresponding to the same selected saliencies. 
Recognizing that HRTFs can be smoothed without affecting perceived sound localization \cite{Kistler92PcaSmooth,kulkarni1998role}, these HRTFs were derived using a PCA-based smoothing approach. Specifically, they were reconstructed from class centroids in a lower-dimensional space retaining 90\% of the variance, ensuring more robust averaged HRTFs.
These class-conditional averaged HRTFs were then fed into the same model to generate predictions and their corresponding saliencies.
Figure~\ref{fig:MEAN_SALIENCY_BANDS_CombinedModel_AGAINST_EACH_DATASET}b) shows these HRTFs as black solid lines and its corresponding saliencies as a colored background as in the previous subfigure.
To illustrate that key spectral features are accurately represented and not obscured by averaging, the HRTF closest to the averaged one from each dataset has been also displayed using different colors and linestyles.

Also for illustration purposes, Figure~\ref{fig:MEAN_SALIENCY_BANDS_CombinedModel_AGAINST_EACH_DATASET}c) shows the particular HRTFs  along with their corresponding saliency for subject 1122 from the BiLi dataset (the one correctly classified with highest confidence).
The fact that specific saliency bands are close to the ones for the averaged HRTF and in the MSC indicates a
%
strong alignment between the subject's HRTF features and the model's learned patterns. The next section offers a detailed discussion of the most relevant findings derived from these experiments.

\begin{figure*}

    \centering
    \includegraphics[width=\textwidth]{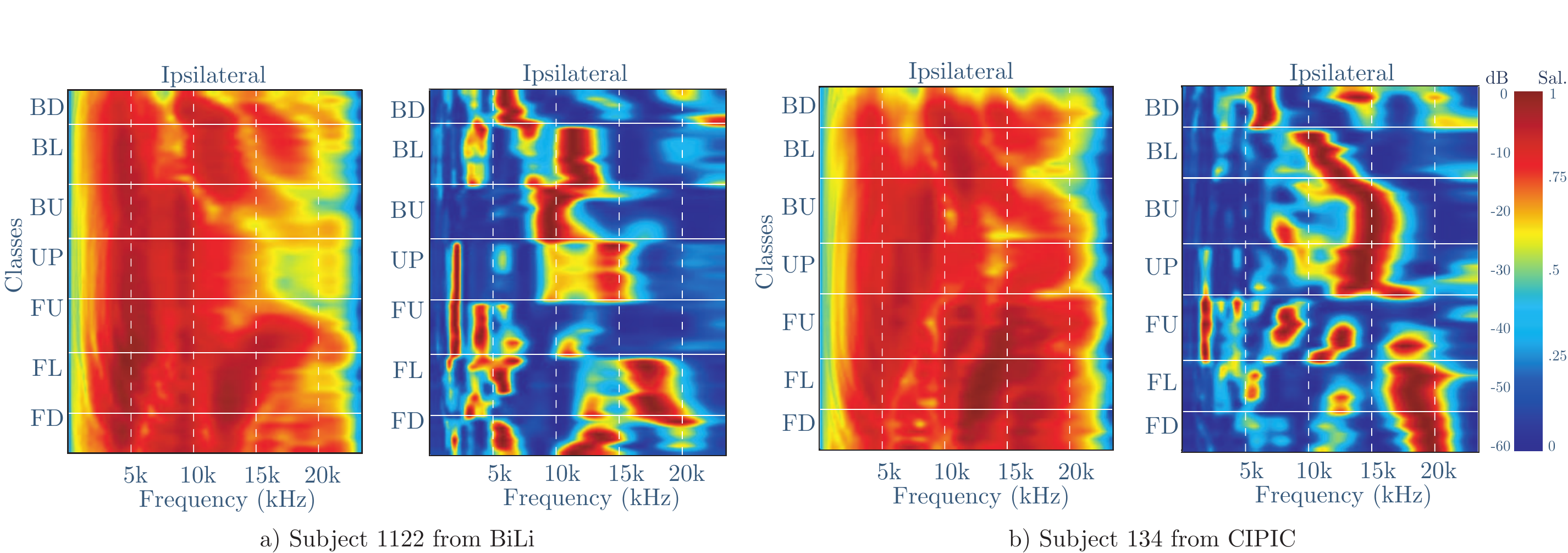}

    \caption{Sagittal HRTFs and their associated saliency maps for two representative subjects chosen based on overall classification confidence.}
    \label{fig:Continuous_Saliency_Contours}
\end{figure*}

\section{Discussion}
\label{sec:discussion}
\subsection{Frontal Hemisphere}

As observed in the first four rows of Figure~\ref{fig:MEAN_SALIENCY_BANDS_CombinedModel_AGAINST_EACH_DATASET}a), our analysis reveals distinct saliency patterns for frontal source locations.  For frontal level angles (FL), high saliency is observed at a high frequency band from 16 to 22 kHz with the highest saliency peak at 18 kHz, followed in contribution by an uniform band spanning lower frequencies, from 2 to 6 kHz with the saliency peak centered around 5 kHz. As source elevation shifted from frontal level either upwards (FU) or downwards (FD), the contribution of low frequencies increased with peaks in bands as low as 2 kHz. Towards forward down (FD) directions the saliency zones become more concise with 4 contribution peaks (ordered by magnitude) at 18 kHz, 12 kHz, 5 kHz and 2 kHz. For both frontal up (FU) and above (UP) angles, the saliency contours exhibited a stronger saliency concentration in the low-frequency range spanning from 1 to 5 kHz exhibiting the main peaks of saliency around 2 kHz. 

The importance of low frequencies in elevation perception, particularly for sound sources located away from the median plane, has been documented in the literature. For example, experiments in \cite{Algazi2001ElevationSpectralCuesLowFreq} demonstrated that subjects could discern elevations from random noise bursts filtered to 3 kHz. Additionally, \cite{zonooz2018learning} provided evidence that listeners deprived of high-frequency information (above 6 kHz) relied more heavily on low-frequency bands for elevation perception.  The band between 4 kHz and 8 kHz has also been shown to be important for elevation, particularly in frontal directions \cite{yao2020role, iwaya2012role, Zonooz2019}.

Our findings indicate that low-frequency regions (below 5 kHz) exhibit a strong influence on the perception of ``front-up" and ``up" elevations. The relevance of very high frequencies (above 10 kHz) appears to increase from zenithal angles (UP), where their saliency contribution is simmilar to those of lower frequencies, to level (FL) and below-the-head (FD) angles where they concentrate the saliency. While for front directions the influence of high-frequency modifications caused by pinna structures (e.g., concha and scapha) may be increased, further investigation is needed to better understand these effects.

Another interesting finding derived from the saliency contours from actual HRTFs can be related to the idea of gradient-based localization \cite{zakarauskas1993computational}. As observed in Figure~\ref{fig:MEAN_SALIENCY_BANDS_CombinedModel_AGAINST_EACH_DATASET}c), saliency bands are predominantly positioned along slopes rather than being centered at peaks and notches. 

To illustrate better all the above observations, Figure~\ref{fig:Continuous_Saliency_Contours} shows the ipsilateral HRTF map along samples taken in the sagittal plane for two representative subjects (i.e., those with the highest prediction confidence) from different datasets. Note that saliency regions cover in general high-frequency zones above 10 kHz with some effects of high saliency concentration in the low-frequency range below 5 kHz for front upper directions.

\subsection{Rear Hemisphere}

As observed in the rows labeled as BU, BL, BD of Figure~7a), corresponding to the rear source locations, some common patterns emerged. Saliency for back-up elevations (BU) is concentrated at the high frequency band from 5 to 18 kHz, with the main peaks at 8 and 15 kHz, showing very low saliency in other frequencies. When shifting downwards the saliencies become more spread, moving towards lower frequency bands. At back level directions (BL), low frequencies become more significant, and the model displays a first peak of saliency around 12 kHz and another one at 4 kHz. However these peaks are less significant and the saliency is more even. Going further down towards back down directions (BD), saliencies shift towards lower frequencies with the strongest saliency peak at 6 kHz. Very low frequencies in the range 200 to 1500 Hz also show slight contributions for the classes BU and BL. 

The significant change in the saliency contour for rear directions suggests that elevation cues might be distributed across a wider frequency range in the back compared to the front. This aligns with the complex and non-monotonic relationship observed between spectral contrast and elevation gain in other studies \cite{Zonooz2019}, highlighting the multifaceted nature of sound localization cues. In a recent work, Zielinski \cite{zielinski2022} 
highlighted the 5 kHz band as relevant for front-back discrimination in binaural recordings, particularly in the horizontal plane. Our findings align with this observation, as we identified a prominent saliency in the 2-6 kHz band for both the frontal-level (FL) and back-level (BL) sectors. Similarly, the frequency range from 400 Hz to 1.2 kHz was also found to be relevant in \cite{yao2020role} based on an $F$-ratio analysis of CIPIC HRTFs. This suggests that low frequencies, while not the dominant cues, may still play a role in conveying elevation information.

\subsection{Lateral angles}

Finally, the saliency profiles for lateral up (LU) and lateral down (LD) angles exhibited slightly different patterns. Both LU and LD exhibit high saliency in very high frequencies between 16 and 22 kHz with a peak at 20 kHz and a low frequency band between 1.5 and 3 kHz with a peak around 2 kHz. However the other saliency bands present a complementary effect: LU exhibits a main peak of saliency centered at 8.5 kHz  (similar to classes FU, UP, BU) and a significant contribution at very low frequencies around 200 to 1000 Hz, while presenting low saliency in the others. In contrast, LD exhibits high saliency between 12 and 14 kHz (similar to FD and BD), with low saliency in the other bands.

\section{Conclusion}
\label{sec:conclusion}
In this study, we employed explainable artificial intelligence techniques, specifically CAM, on a baseline CNN model trained on a wide variety of public HRTF datasets. Our aim was to identify common frequency bands across datasets that are meaningful for classifying HRTFs into different elevation sectors, irrespective of the different effects produced by varying measurement settings and minimizing subject-dependent cues. To this end, we also explored different pre-processing techniques suitable for standardizing HRTF data.
Regarding the standardization methods, we found that the best inter-dataset results were obtained using  linear magnitude normalization based on Average Equator Energy. Additionally, adding ERB-based filtering improved generalization capabilities while maintaining similar classification results. Moreover, it was determined that the best results are obtained through data diversity in training, where a small percentage of samples from each dataset is considered for building a combined model. 
Finally, we utilized this model to identify common relevant bands by analyzing prediction saliencies across all the different datasets for each particular elevation sector. Our analysis revealed distinct saliency patterns across different source elevation angles. Generally the saliency was found to be concentrated at high frequencies over 5 kHz, pointing that the main spectral cues for elevation may be found at these bands. However low frequency bands between 1 and 5 kHz with a peak at 2 kHz also have an important contribution for discriminating elevation from front and lateral directions. For upward directions,
high common saliency was found between 10 and 12 kHz. Subject-specific results revealed also that elevation cues tend to be aligned with prominent negative and positive slopes in the vecinity of peaks and notches, which supports the idea of gradient-based localization cues. These results are considerably consistent with previous experimental findings in the literature.
While this study provides initial insights into the primary elevation cues leveraged by neural networks for HRTF-based elevation classification, there are several areas for future exploration. One key direction involves performing listening tests to explore how the presented saliency analysis correlates to human auditory perception. Future work could also explore the impact of different neural architectures and more sophisticated XAI techniques to assess whether alternative methods offer deeper or more intuitive insights.

\section*{Acknowledgment}
This work has been supported by Grants FPU20/05384 from the  Ministry of Universities of Spain, RYC2020-030679-I from MCIN/AEI/-10.13039/501100011033, as well as by the ``European Social Fund (ESF) Investing in Your Future". Grant TED2021-131003B-C21 funded by MCIN/AEI/10.13039/501100011033 and by the “EU NextGeneration EU/PRTR”. Grant PID2022-137048OB-C41 funded by MICIU/AEI/-10.13039/501100011033 and “ERDF A way of making Europe”.
The authors acknowledge also the Artemisa computer resources funded by the EU ERDF and Comunitat Valenciana, and the technical support of IFIC (CSIC-UV).

\appendix

\bibliographystyle{elsarticle-num} 
\bibliography{Main-elsarticle-num}






\end{document}